\documentclass[%
 reprint,
 amsmath,amssymb,
 aps,
]{revtex4-2}

\usepackage{ytableau}
\usepackage[hyperfootnotes=false]{hyperref}
\hypersetup{
 colorlinks=true,
 citecolor=blue,
 linkcolor=red,
 urlcolor=blue}
\usepackage{mathtools}
\usepackage{nccmath} 
\usepackage{xcolor}
\usepackage{caption}
\usepackage{subcaption}
\usepackage{graphicx}% Include figure files
\usepackage{dcolumn,physics}% Align table columns on decimal point
\usepackage{tabularx}
\usepackage{bm}% bold math
\begin{document}

\preprint{APS/123-QED}

\title{Spectroscopic Analysis of Fully Heavy Pentaquarks}% Force line breaks with \\
\author{Rashmi, Alka Upadhyay}
 \affiliation{Department of Physics and Material Science, Thapar Institute of Engineering and Technology, Patiala, India, 147004}
 
\email{gargrshmiphy@gmail.coma}

\date{\today}% It is always \today, today,

\begin{abstract}

Motivated by the discovery of the fully charmed tetraquark state $X(6900)$ in the invariant mass spectrum of $J/\psi$ pairs by the LHCb collaboration, this study explores the potential existence of fully heavy pentaquark states. We systematically investigate the low-lying s-wave fully heavy pentaquark states across all possible configurations.
The classification of these states is performed using the Young-Yamounachi bases through the Young-tableau technique. We analyze the mass spectrum and magnetic moments of pentaquarks with quantum numbers $J^P$= $\frac{1}{2}^{\pm}$,$\frac{3}{2}^{\pm}$ and $\frac{5}{2}^{\pm}$ 
utilizing effective mass and screened charge schemes. Our findings are compared with various theoretical models, providing valuable insights for future experimental studies.
\end{abstract}

\maketitle

%\tableofcontents

\section{Introduction}
Over the past decades, several exotic hadrons have been discovered at LHCb and other experimental facilities. In 2020, a fully-charmed X(6900) tetraquark structure with minimal quark content $c\Bar{c}c\Bar{c}$ had been reported by LHCb collaboration in the $J/\psi$ pair invariant mass spectrum with a statistical significance of more than 5 $\sigma$ \cite{X(6900)}. The mass and width of the X(6900) resonance are measured to be:
\begin{equation*}
    M = 6886 \pm 11 \pm 11 MeV, \hspace{0.3cm}
    \Gamma = 168 \pm 33 \pm 69 MeV
\end{equation*}
Many theoretical frameworks have been proposed to study this fully charmed tetraquark state, such as QCD sum rules \cite{WangQCD}, string junction picture \cite{String}, the potential model \cite{liu}, the non-relativistic diquark-antidiquark model \cite{lundhammar}, and the constituent quark model \cite{Jin}. Since such states with very large energies may be accessible experimentally and easily separated from other states, the finding of a completely charm tetraquark state received a lot of attention. In 2022,LHcb collaboration discoverd a hidden charm pentaquark $P_{c}$ (4380) with minimal quark content $udsc\Bar{c}$, observed in the $J/\psi$ mass spectrum from $B^- \rightarrow J/\psi\Lambda p$ decays with the statistical significance of 15 $\sigma$ \cite{chen2022}. The mass and the width of the new pentaquark are measured to be 4338.2 ± 0.7 ± 0.4MeV and 7.0 ± 1.2 ± 1.3 MeV, respectively. The preferred quantum numbers are $J^P$ = 1/2. Also, in 2021, the pentaquark state $P_{cs}$(4459) was observed in the $J/\psi\Lambda$ invariant mass distribution from an amplitude analysis of the $\Xi_{b}^- \rightarrow J/\psi \Lambda K^-$ decay processes \cite{2021}. The observed structure is consistent with a hidden-charm pentaquark with strangeness, characterized by a mass of $4458.8 \pm 2.9_{-1.1}^{+4.7}$ MeV and a width of $17.3 \pm 6.5_{-5.7}^{+8.0}$ MeV. The spin is expected to be 1/2 or 3/2, and its parity can be either $\pm$ 1. Furthermore, in 2019, LHCb reported  the discovery of three hidden-charm pentaquark
structures, $P_{c}$(4312) decaying to $J/\psi p$, with a statistical significance of 7.3 $\sigma$ in a data sample of $\Lambda^{0}_{b}$ $\rightarrow$ $J/\psi pK^-$ decays and $P_{c}$(4450) pentaquark structure consisting of two narrow overlapping peaks, $P_{c}$(4440) and $P_{c}$(4457), with a statistical significance of 5.4 $\sigma$ forthe two-peak interpretation is 5.4 $\sigma$ \cite{2019}. Also, in 2015, the LHCb collaboration reported two hidden-charm structures, $P_{c}$(4380), and $P_{c}$(4450), in $\Lambda_{b}$ decay \cite{2015}. These
two pentaquark states have a mass $4380 \pm  8 \pm  28$ MeV
and $4449.8 \pm  1.7 \pm 2.5$ MeV with corresponding widths
  of $205 \pm  18 \pm  86$ MeV and $39 \pm  5 \pm  19$ MeV respectively. The discovery of  X(6900) tetraquark and $P_{c}$ states has sparked the keen interest about the possible existence of fully heavy pentaquark states $QQQQ\bar{Q}$ where $Q = c,b$. Using QCD sum rules approach, the energies of bound states $cccc\Bar{b}$ and $bbbb\Bar{c}$ with $J^P$ =$\frac{1}{2}$ and $J^P$ =$\frac{3}{2}$ have been studied. Furthermore, a comprehensive analysis of the mass spectra of s-wave fully heavy pentaquark states $QQQQ\Bar{Q}$ have been conducted using various models such as the chromomagnetic interaction (CMI) model, constituent quark model with variational method, lattice-QCD inspired quark model with complex scaling method, quark models with resonating group method, and MIT bag model. In reference, the author systematically explored not only fully charmed(bottom) pentaquarks with a color-singlet structure. We perform a theoretical investigation of fully heavy pentaquark states in effective mass and screened charge models \cite{VERMA}. Many theoretical models have been proposed to study fully heavy pentaquark states such as the MIT bag model \cite{MIT}, quark model \cite{QuarkM}, chromomagnetic interaction model \cite{chromo}, QCD sum rules \cite{QCD, QCD1}, Constituent quark model \cite{CQM} etc. for the understanding of these exotic particles. In the work of reference \cite{sharma}, the study of hidden-charm pentaquarks is done using an effective mass and screened charge scheme. Moreover, hidden-bottom and singly heavy pentaquarks with quantum numbers $J^P$= $\frac{1}{2}^{\pm}$,$\frac{3}{2}^{\pm}$ and $\frac{5}{2}^{\pm}$ investigated in the references\cite{sharma1,sharma2,sharma20241}. Similarly, we estimated the mass spectrum and magnetic moments of the pentaquark system in the framework of effective mass and screened charge. The mass spectrum alone cannot reveal their internal structure; therefore, the magnetic moment plays a significant role in understanding these multiquark states' inner structure and dynamics. The pentaquark's magnetic moment encoding provides important details about the charge and magnetization distributions inside hadrons, which is helpful for analyzing their geometrical shapes. This study is significant because magnetic moments provide insight into the internal structure of the pentaquark. Magnetic moments and transition moments of baryons have been studied in effective mass and screened charge schemes \cite{dhir}. So, we aim to investigate the magnetic moments of fully heavy pentaquarks in a similar manner.\\
This work is organized as follows. In section II, the classification of fully heavy pentaquarks with different configurations has been done using the Young Tableau technique. Section III introduces the methodology and describes how the masses and magnetic moments of these fully heavy pentaquarks are computed. Section IV presents the results and discusses their implications. In section V, a brief review of the work is given.

\section{Theoretical Framework}
To classify the fully heavy pentaquark states, we employed the group theory approach to construct their wave functions. For this, certain rules must be followed. Firstly, these states must be a color singlet. The wave function of the pentaquarks should be antisymmetric according to the Pauli exclusion principle. Both color confinement and the Pauli exclusion principle are essential in describing the symmetry of the wave function. Further, we used Young-Yamanouchi bases to construct the pentaquark wavefunction via the Young Tableaux technique \cite{zhang}. After that, we can use these pentaquark wave functions to calculate the mass spectra and magnetic moments of the corresponding pentaquark
states. The wave function of multiquark states contains spatial, color, spin, and flavor degrees of freedom. Therefore, the total wave function of the pentaquark system can be defined as:
\begin{equation}
    \psi_{tot} =\phi_{spatial}\otimes\eta_{flavor}\otimes\zeta_{color}\otimes\chi_{spin}
\end{equation}
We only consider low-lying s-wave pentaquark states, thus the symmetrical constraint from the spatial pentaquark wave function is minimal. While exchanging identical quarks in different configurations of a pentaquark system, their $\eta_{flavor}\otimes\zeta_{color}\otimes\chi_{spin}$ wave function must be antisymmetric in nature.
In the group theory approach, each quark is assigned as a fundamental representation $SU(n)$ and antiquark by $SU(n)$ where $n=2,3,4,5,...$ for spin, flavor, color, and spin-flavor degrees of freedom, respectively. The corresponding algebraic structure
consists of usual spin-flavor and color algebras:
\begin{align*}
    SU_{SF}(6) \otimes SU_{C}(3)
\end{align*}       
with
\begin{equation}
     SU_{SF}(6) = SU_{F}(3) \otimes SU_{S}(2)
\end{equation}
In color space algebra, due to the color confinement property, we consider those pentaquark ($Q_1 Q_2 Q_3 Q_4\bar{Q}_5$) configurations that are color singlet in nature. Therefore, the first four quarks of the pentaquark system in equation (4) are in color triplet and can be expressed in the form of Young Tableau as follows \cite{Hong}:
\begin{align*}
   \begin{ytableau}
    \none  & 1 & 2  \\
    \none  & 3 \\
    \none  & 4
\end{ytableau}\hspace{0.8cm}
\begin{ytableau}
    \none & 1 & 3 \\
    \none & 2 \\
    \none & 4
\end{ytableau}\hspace{0.8cm}
\begin{ytableau}
    \none & 1 & 4 \\
    \none & 2 \\
    \none & 3
\end{ytableau}
\end{align*}

When the antitriplet from the antiquark $\bar{5}$ is paired with the three color triplets listed above, it gives three color singlet configurations of a fully heavy pentaquark system.
\\
In the flavor space, we classify the  
pentaquark system into three subsystems based on their symmetries by defining their flavor states in terms of $\alpha$, $\beta$ : (1) where the first four quarks are identical, such as $\alpha\alpha\alpha\alpha\bar{\beta}$, $\beta\beta\beta\beta\bar{\alpha}$ (2) where the first three quarks are identical, includes
$\alpha\alpha\alpha\beta\bar{\beta}$
or
$\beta\beta\beta\alpha\bar{\alpha}$
pentaquark subsystems, and (3) involving two pairs of identical quarks, such as
$\alpha\alpha\beta\beta\bar{\beta}$,
 $\beta\beta\alpha\alpha\bar{\alpha}$
pentaquark subsystems where $\alpha, \beta$ = $c,b$.
\\
Similarly, we can define the spin wave functions by taking the product of five fermions in terms of Young Tableau as given below \cite{MIT}:
\begin{table}[h]
\centering
    \caption{Young Tableau representation for  Pentaquarks}
   
      \begin{tabular}{ccc}
      
     \begin{ytableau}
     \none &
\end{ytableau}\hspace{0.3cm}$\otimes$
\begin{ytableau}
     \none &
\end{ytableau}  $\otimes$
\begin{ytableau}
     \none &
\end{ytableau}\hspace{0.2cm} $\otimes$
 \begin{ytableau}
     \none &
\end{ytableau} \hspace{0.2cm}$\otimes$
\begin{ytableau}
     \none &
\end{ytableau} = \nonumber \\
\\
\begin{ytableau}
    \none &  &  &  &  &
\end{ytableau} \hspace{0.1cm}$\oplus$
 4\begin{ytableau}
    \none &  &  &  &  \\
    \none &
\end{ytableau}\hspace{0.1cm} $\oplus$
5\begin{ytableau}
    \none &  &  & \\
    \none &  &
\end{ytableau}\\
    \end{tabular}

\end{table}
\\
we obtained various configurations of spin wave functions in terms of Young-Yamanouchi bases of one, four, and five dimensions corresponding to their spin multiplets\cite{zhang2}.
\\
For $J^P$ = 5/2, there is only $\chi_1$ symmetry as given below:
\\

\begin{ytableau}
             \none & 1 & 2 & 3 & 4 & 5
\end{ytableau}

$\chi^P_{1}$\\
Similarly, for $J^P$ = 3/2, there are four combinations of spin wave functions. These wave functions can be written as:
\\

\begin{ytableau}
    \none & 1 & 2 & 3  & 4 \\
    \none & 5
\end{ytableau} $\chi^P_{2}$\hspace{0.3cm}
\begin{ytableau}
    \none & 1 & 2 & 3  & 5 \\
    \none & 4
\end{ytableau} $\chi^P_{3}$\hspace{0.3cm}
\\
\\

\begin{ytableau}
    \none & 1 & 2 & 5  & 4 \\
    \none & 3
\end{ytableau} $\chi^P_{3}$\hspace{0.3cm}
\begin{ytableau}
    \none & 1 & 3 & 4  & 5 \\
    \none & 2
\end{ytableau} $\chi^P_{4}$\hspace{0.3cm}
\\
\\
and for the case of $J^P$ =1/2,
\\
\\
\begin{ytableau}
    \none & 1 & 2 & 3 \\
    \none & 4 & 5
\end{ytableau}
$\chi^P_{6}$
    \begin{ytableau}
    \none & 1 & 2 & 4 \\
    \none & 3 & 5
\end{ytableau}
$\chi^P_{7}$
\begin{ytableau}
    \none & 1 & 3 & 4 \\
    \none & 2 & 5
\end{ytableau}$\chi^P_{8}$
\\
\\
\\
\begin{ytableau}
    \none & 1 & 2 & 5 \\
    \none & 3 & 4
\end{ytableau}
$\chi^P_{9}$
\begin{ytableau}
    \none & 1 & 3 & 5 \\
    \none & 2 & 4
\end{ytableau}
$\chi^P_{10}$
\\

In this work, there are a total of ten symmetries available from $\chi_1$ to $\chi_{10}$ ranging from  spin 5/2 to spin 1/2. We utilized the $\chi_1$, $\chi_4$, and $\chi_8$ symmetries for spin wave functions of 5/2, 3/2, and 1/2, respectively, for the computation of magnetic moments of fully heavy pentaquarks in the framework of effective mass and screened charge described in subsections A and B. We used $\chi_1$, $\chi_4$ and $\chi_8$ for spin 5/2, 3/2 and 1/2 respectively.
\\

%%%%%%%%%%%%%%%%%
 \begin{table*}[ht!]
 \centering
\caption{Mass spectrum of ground state fully heavy pentaquarks for $J^P$ = 1/2, 3/2, and 5/2 using effective mass scheme. Masses are in the unit of MeV. }
\begin{tabular}{cccccccccc}
       \hline
       \hline
Quark Content & \hspace{0.3cm} S &  \hspace{0.3cm} Our Prediction &\hspace{0.3cm} Ref. \cite{QuarkM} & \hspace{0.3cm} Ref\cite{QuarkM}  & \hspace{0.3cm} Ref. \cite{MIT} & \hspace{0.3cm} Ref. \cite{CQM} & \hspace{0.3cm} Ref. \cite{Gang}\\
         \hline
         \\
      $cccc\Bar{c}$  & \hspace{0.3cm} 1/2  & \hspace{0.3cm} 8537.4 & \hspace{0.3cm} 8538.6 & \hspace{0.3cm} 8428.9 & \hspace{0.3cm} 8229 & \hspace{0.3cm} 8193.2 & \hspace{0.3cm} 8144\\
      \\
        $cccc\Bar{c}$ & \hspace{0.3cm} 3/2  & \hspace{0.3cm} 8547.4 & \hspace{0.3cm} 8536.8  & \hspace{0.3cm} 8426.8 & \hspace{0.3cm} 8262 & \hspace{0.3cm} 8144.6  & \hspace{0.3cm} 8155\\
        \\
          $cccc\Bar{c}$ & \hspace{0.3cm} 5/2  &  \hspace{0.3cm} 8562.5& \hspace{0.3cm} 8538.6 & \hspace{0.3cm} 8429.8 & \hspace{0.3cm}$\cdots$ & \hspace{0.3cm}$\cdots$ & \hspace{0.3cm} $\cdots$\\
          \\
           \hline
           
         \\
      $bccc\Bar{c}$  & \hspace{0.3cm} 1/2 & \hspace{0.3cm} 11873.2 & \hspace{0.3cm} $\cdots$ & \hspace{0.3cm} $\cdots$ & \hspace{0.3cm} 11526 & \hspace{0.3cm} 11438.2  & \hspace{0.3cm}$\cdots$ \\
      \\
        $bccc\Bar{c}$ & \hspace{0.3cm} 3/2 & \hspace{0.3cm} 11880.3  & \hspace{0.3cm} $\cdots$ & \hspace{0.3cm} $\cdots$ & \hspace{0.3cm}11549 & \hspace{0.3cm} 11443.7 & \hspace{0.3cm}$\cdots$ \\
        \\
          $bccc\Bar{c}$ & \hspace{0.3cm} 5/2 &  \hspace{0.3cm} 11891.3  & \hspace{0.3cm} $\cdots$ & \hspace{0.3cm} $\cdots$ & \hspace{0.3cm}11554 & \hspace{0.3cm}$\cdots$ & \hspace{0.3cm}$\cdots$  \\
          \\
           \hline
         \\
      $bbcc\Bar{c}$  & \hspace{0.3cm} 1/2 & \hspace{0.3cm} 15208.8  & \hspace{0.3cm} $\cdots$  & \hspace{0.3cm} $\cdots$ & \hspace{0.3cm} 14852 & \hspace{0.3cm} 14566.0  & \hspace{0.3cm}$\cdots$ \\
      \\
        $bbcc\Bar{c}$ & \hspace{0.3cm} 3/2  & \hspace{0.3cm} 15212.8 & \hspace{0.3cm} $\cdots$  & \hspace{0.3cm} $\cdots$ & \hspace{0.3cm} 14866 & \hspace{0.3cm} 14579.6 &  \hspace{0.3cm}$\cdots$ \\
        \\
          $bbcc\Bar{c}$ & \hspace{0.3cm} 5/2 & \hspace{0.3cm} 15221.8 & \hspace{0.3cm} $\cdots$  & \hspace{0.3cm} $\cdots$ & \hspace{0.3cm}14872 & \hspace{0.3cm}$\cdots$ &  \hspace{0.3cm}$\cdots$ \\
          \\
           \hline  
         \\
      $bbbc\Bar{c}$  & \hspace{0.3cm} 1/2 & \hspace{0.3cm} 18543.2  & \hspace{0.3cm} $\cdots$  & \hspace{0.3cm} $\cdots$ & \hspace{0.3cm} 18159& \hspace{0.3cm} 17883.8  &  \hspace{0.3cm} $\cdots$ \\
      \\
        $bbbc\Bar{c}$ & \hspace{0.3cm} 3/2& \hspace{0.3cm} 18546.3 & \hspace{0.3cm} $\cdots$  & \hspace{0.3cm} $\cdots$ & \hspace{0.3cm} 18171& \hspace{0.3cm} 17891.2  &  \hspace{0.3cm}$\cdots$ \\
        \\
          $bbbc\Bar{c}$ & \hspace{0.3cm} 5/2 & \hspace{0.3cm} 18552.8 & \hspace{0.3cm} $\cdots$  & \hspace{0.3cm} $\cdots$ & \hspace{0.3cm}18183 & \hspace{0.3cm}$\cdots$ & \hspace{0.3cm} $\cdots$\\
          \\
           \hline
           
         \\
      $bbbb\Bar{c}$  & \hspace{0.3cm} 1/2 & \hspace{0.3cm} 21879.7 & \hspace{0.3cm} $\cdots$  & \hspace{0.3cm} $\cdots$ & \hspace{0.3cm} 21491 & \hspace{0.3cm} 21025.6 & \hspace{0.3cm} $\cdots$ \\
      \\
        $bbbb\Bar{c}$ & \hspace{0.3cm} 3/2 & \hspace{0.3cm} 21879.0  & \hspace{0.3cm} $\cdots$  & \hspace{0.3cm} $\cdots$ & \hspace{0.3cm} 21472 & \hspace{0.3cm} 20794.5 &  \hspace{0.3cm}$\cdots$ \\
        \\
          $bbbb\Bar{c}$ & \hspace{0.3cm} 5/2 &  \hspace{0.3cm} 21884.3 & \hspace{0.3cm} $\cdots$  & \hspace{0.3cm} $\cdots$ & \hspace{0.3cm}$\cdots$ & \hspace{0.3cm}$\cdots$&\hspace{0.3cm} $\cdots$ \\
          \\
           \hline
          \\
          $cccc\Bar{b}$ & \hspace{0.3cm} 1/2 &  \hspace{0.3cm} 11867.7 & \hspace{0.3cm} $\cdots$  & \hspace{0.3cm} $\cdots$ & \hspace{0.3cm} 11582 & \hspace{0.3cm} 11501.5   & \hspace{0.3cm} $\cdots$  \\
          \\
           $cccc\Bar{b}$ & \hspace{0.3cm} 3/2  & \hspace{0.3cm} 11887.6 & \hspace{0.3cm} $\cdots$  & \hspace{0.3cm} $\cdots$ & \hspace{0.3cm} 11569 & \hspace{0.3cm} 11477.8 & \hspace{0.3cm} $\cdots$\\
          \\
            $cccc\Bar{b}$ & \hspace{0.3cm} 5/2  & \hspace{0.3cm} 11891.3 & \hspace{0.3cm} $\cdots$  & \hspace{0.3cm} $\cdots$ & \hspace{0.3cm}$\cdots$& \hspace{0.3cm}$\cdots$  & \hspace{0.3cm} $\cdots$\\
            \\
       \hline
       
          \\
          $cccb\Bar{b}$ & \hspace{0.3cm} 1/2  &   \hspace{0.3cm} 15214.7 & \hspace{0.3cm} $\cdots$  & \hspace{0.3cm} $\cdots$ & \hspace{0.3cm} 14862 & \hspace{0.3cm} 14676.3  &  \hspace{0.3cm} $\cdots$ \\
          \\
           $cccb\Bar{b}$ & \hspace{0.3cm} 3/2  & \hspace{0.3cm} 15218.3 & \hspace{0.3cm} $\cdots$  & \hspace{0.3cm} $\cdots$ & \hspace{0.3cm}14862 & \hspace{0.3cm} 14687.2  & \hspace{0.3cm} $\cdots$\\
          \\
            $cccb\Bar{b}$ & \hspace{0.3cm} 5/2 & \hspace{0.3cm} 15221.8 & \hspace{0.3cm} $\cdots$  & \hspace{0.3cm} $\cdots$  & \hspace{0.3cm} 14873& \hspace{0.3cm}$\cdots$ & \hspace{0.3cm} $\cdots$ \\
            \\
         \hline
          \\
          $ccbb\Bar{b}$ & \hspace{0.3cm} 1/2 &   \hspace{0.3cm} 18546.1 & \hspace{0.3cm} $\cdots$  & \hspace{0.3cm} $\cdots$ & \hspace{0.3cm} 18154& \hspace{0.3cm} 17784.5  & \hspace{0.3cm} $\cdots$ \\
          \\
           $ccbb\Bar{b}$ & \hspace{0.3cm} 3/2 & \hspace{0.3cm} 18549.8 & \hspace{0.3cm} $\cdots$  & \hspace{0.3cm} $\cdots$ & \hspace{0.3cm} 18164 & \hspace{0.3cm} 17784.9  &   \hspace{0.3cm} $\cdots$\\
          \\
            $ccbb\Bar{b}$ & \hspace{0.3cm} 5/2 & \hspace{0.3cm} 18552.8 & \hspace{0.3cm} $\cdots$  & \hspace{0.3cm} $\cdots$ & \hspace{0.3cm} 18182& \hspace{0.3cm}$\cdots$ & \hspace{0.3cm} $\cdots$ \\
             \hline
          \\
          $cbbb\Bar{b}$ & \hspace{0.3cm} 1/2 &   \hspace{0.3cm} 21878.9 & \hspace{0.3cm} $\cdots$  & \hspace{0.3cm} $\cdots$ & \hspace{0.3cm} 21472 & \hspace{0.3cm} 21079.0 & \hspace{0.3cm} $\cdots$ \\
          \\
           $cbbb\Bar{b}$ & \hspace{0.3cm} 3/2  & \hspace{0.3cm} 21882.8 & \hspace{0.3cm} $\cdots$  & \hspace{0.3cm} $\cdots$ & \hspace{0.3cm} 21475 & \hspace{0.3cm} 21091.6  &  \hspace{0.3cm} $\cdots$\\
          \\
            $cbbb\Bar{b}$ & \hspace{0.3cm} 5/2 & \hspace{0.3cm} 21884.3 & \hspace{0.3cm} $\cdots$  & \hspace{0.3cm} $\cdots$ & \hspace{0.3cm} 21480 & \hspace{0.3cm}$\cdots$ & \hspace{0.3cm} $\cdots$ \\
            \\
       
       \hline
          \\
          $bbbb\Bar{b}$ & \hspace{0.3cm} 1/2 &   \hspace{0.3cm} 25212.8 & \hspace{0.3cm} 25275.6 & \hspace{0.3cm} 25179.4 & \hspace{0.3cm} 24761 & \hspace{0.3cm} 24248.0  & \hspace{0.3cm} 24443 \\
          \\
           $bbbb\Bar{b}$ & \hspace{0.3cm} 3/2 & \hspace{0.3cm} 25214.8 & \hspace{0.3cm} 25275.7  & \hspace{0.3cm} 25179.4 & \hspace{0.3cm} 24770 & \hspace{0.3cm} 24210.7  &  \hspace{0.3cm} 24374 \\
          \\
            $bbbb\Bar{b}$ & \hspace{0.3cm} 5/2 & \hspace{0.3cm} 25216.3  & \hspace{0.3cm} 25275.9 & \hspace{0.3cm} 25179.2 & \hspace{0.3cm} 24302 & \hspace{0.3cm}$\cdots$ & \hspace{0.3cm} 24302  \\
            \\
       \hline
       \hline
       \end{tabular}
        \label{tab:2}
   \end{table*}

\subsection{Effective quark mass scheme}
We computed the effective mass of quarks (antiquarks) by considering their interaction with neighboring quarks through one gluon exchange scheme. By using effective quark masses, we calculated the magnetic moments of fully heavy pentaquarks, which helped us to explore their inner structure. The mass of pentaquarks can be written as the sum of quark masses plus spin-dependent hyperfine interaction term \cite{kumar}  :
\begin{equation}
    M_P = \sum_{i=1}^5 m_i^{eff} = \sum_{i=1}^5 m_i + \sum_{i<j} b_{ij} s_i.s_j
\end{equation}
 here, $s_i$ and $s_j$ represent the spin operator for the $i^{th}$ and $j^{th}$ quarks (antiquark) and $m_i^{eff}$ represents the effective mass for each of the quark (antiquark) and $b_{ij}$ is defined as:
 \begin{equation}
     b_{ij} = \frac{16 \pi \alpha_s}{9 m_i m_j} \bra{\Psi_0}\delta^3(\Vec{r})\ket{\Psi_0}
 \end{equation}
for the pentaquarks, where $\Psi_0$ is the pentaquark function. Additionally, a spin-independent interaction term may exist, which can be modeled by adjusting quark masses through renormalization. Consequently, the interaction with other quarks may result in a modification of the mass of the quark within the pentaquark $P(12345)$. For ($aaaab$) pentaquarks, where four quarks are identical, we write:
 \begin{equation}
     m^{eff}_1 = m^{eff}_2 = m + \alpha{b_{12}} + \beta{b_{13}} + \gamma{b_{14}} + \eta{b_{15}}
 \end{equation}
 \begin{equation}
     m^{eff}_3 = m^{eff}_4 =  m + \alpha{b_{12}}+\beta{b_{13}}+\gamma{b_{14}}+\eta{b_{15}}
 \end{equation}
 \begin{equation}
     m^{eff}_5 = m_5+ 4\eta{b_{15}}
 \end{equation}
 where,  
 \begin{equation}
     m_1 = m_2 = m_3 = m_4 = m
 \end{equation}
 and
   \begin{equation}
       b_{15} = b_{25} = b_{35} = b_{45} = b_{15}
   \end{equation}
The parameters $\alpha$, $\beta$, $\gamma$ and $\eta$ are calculated from:
\begin{equation}
    M_P = \sum_{i=1}^5 m_i^{eff} = \sum_{i=1}^5 m_i + \sum_{i<j} b_{ij} s_i.s_j
\end{equation}
\begin{equation}
=  4m + m_{5}+\frac{b_{12}}{2} + \frac{b_{13}}{2} + \frac{b_{14}}{2} + b_{15}
\end{equation}
 yielding
 \begin{equation}
     \alpha = \beta = \gamma = \eta = \frac{1}{8}
 \end{equation}
 Therefore,
 \begin{equation}
     m^{eff}_1 = m^{eff}_2 = m + \frac{b_{12}}{8} + \frac{b_{13}}{8} + \frac{b_{14}}{8} + \frac{b_{15}}{8}
 \end{equation}
 \begin{equation}
     m^{eff}_3 = m^{eff}_4 =  m + \frac{b_{12}}{8} +\frac{b_{13}}{8} +\frac{b_{14}}{8} +\frac{b_{15}}{8}
 \end{equation}
 \begin{equation}
     m^{eff}_5 = m_5+ \frac{b_{15}}{2}
 \end{equation}
  In general, for different quarks inside the pentaquark, effective masses equations can be written as \cite{sharma}:
 \begin{equation}
  m_1^{eff} = m_1 + \alpha b_{12} + \beta b_{13} + \gamma b_{14} + \eta b_{15}
 \end{equation}
 
 \begin{equation}
  m_2^{eff} = m_2 + \alpha b_{12} + \beta^{'} b_{23} + \gamma^{'} b_{24} + \eta^{'} b_{25}
  \end{equation}

   \begin{equation}
  m_3^{eff} = m_3 + \beta b_{13} + \beta^{'} b_{23} + \gamma^{''} b_{34} + \eta^{''} b_{35}
   \end{equation}

   \begin{equation}
  m_4^{eff} = m_4 + \gamma b_{14} + \gamma^{'} b_{24} + \gamma^{''} b_{34} + \eta^{'''} b_{45}
 \end{equation}

  \begin{equation}
  m_5^{eff} = m_5 + \eta b_{15} + \eta^{'} b_{24} + \eta^{''} b_{34} + \eta^{'''} b_{45}
 \end{equation}
  here, 1, 2, 3, 4, and 5 stand for $u$, $d$, $s$, $c$, and $b$ quarks. These equations get modified if we consider two/three/four/five identical quarks. Firstly, we are starting with the case of spin-5/2 ($\uparrow\uparrow\uparrow\uparrow\uparrow$), we have \cite{Rohit}:
\begin{equation}
s_1.s_2 = s_2.s_3 =  s_3.s_4 =
 s_4.s_5 = 1/4
\end{equation}
 and parameters are:
 \begin{equation}
     \alpha = \beta = \gamma = \eta = 1/8
 \end{equation}

\begin{equation}
\beta^{'} = \gamma^{'} = \eta^{'} = 1/8
\end{equation}

\begin{equation}
\gamma^{''} = \eta^{''} = \eta^{'''} = 1/8
\end{equation}
therefore, a modified form of the equation of effective mass for $J^P = 5/2^-$ can be written as:
\begin{align}
M_{P_{{5/2}^-}} = m_1 + m_2 + m_3 + m_4 + m_5 + \frac{b_{12}}{4}  + \frac{b_{13}}{4}  + \frac{b_{14}}{4} \nonumber \\
 + \frac{b_{15}}{4}  + \frac{b_{23}}{4}  + \frac{b_{24}}{4}  + \frac{b_{25}}{4}  + \frac{b_{34}}{4}  + \frac{b_{35}}{4}  + \frac{b_{45}}{4}
\end{align}
In a similar manner, for $J^P = 3/2^-$ ($\uparrow\uparrow\uparrow\uparrow\downarrow$),
\begin{equation}
s_1.s_2 =  s_2.s_3 =  s_3.s_4 = 1/4,
\hspace{0.3cm} s_4.s_5 = -1/2
\end{equation}
and total mass of pentaquark for this case is defined as:
\begin{align}
M_{P_{{3/2}^-}} = m_1 + m_2 + m_3 + m_4 + m_5 + \frac{b_{12}}{4}  + \frac{b_{13}}{4}  + \frac{b_{14}}{4} \nonumber \\
 - \frac{b_{15}}{2}  + \frac{b_{23}}{4}  + \frac{b_{24}}{4}  - \frac{b_{25}}{2}  + \frac{b_{34}}{4}  - \frac{b_{35}}{2}  - \frac{b_{45}}{2}
\end{align}
Similarly, we can write for the case of $J^P = 1/2^-$ ($\uparrow\uparrow\uparrow\downarrow\downarrow$) as given below:
\begin{equation}
    s_1.s_2 =s_1.s_3 = s_2.s_3  = 1/4,
\end{equation}
  \begin{equation}
      \hspace{0.3cm} s_1.s_4 = s_1.s_5 = s_2.s_4 = s_2.s_5 = s_3.s_4 = s_3.s_5 = -1/2  
  \end{equation}
\begin{equation}
\hspace{0.3cm} s_4.s_5 = -1/4  
\end{equation}
Therefore, the total mass of the pentaquark for spin 1/2 system can be wrtitten as:
\begin{align}
M_{P_{{1/2}^-}} = m_1 + m_2 + m_3 + m_4 + m_5 + \frac{b_{12}}{4}  + \frac{b_{13}}{4}  - \frac{b_{14}}{2} \nonumber \\
 - \frac{b_{15}}{2}  + \frac{b_{23}}{4}  - \frac{b_{24}}{2}  - \frac{b_{25}}{2}  - \frac{b_{34}}{2}  - \frac{b_{35}}{2}  - \frac{b_{45}}{4}
\end{align}
 The values of the quark masses are taken from Ref. \cite{Rohit} and hyperfine interaction terms $b_{ij}$ are calculated using the effective mass equations.

\begin{align}
     m_u = m_d = \hspace{0.3cm} 362 MeV, \hspace{0.3cm} m_s = \hspace{0.3cm} 539 MeV \nonumber \\
    m_c = \hspace{0.3cm} 1710 MeV, \hspace{0.3cm} m_b = \hspace{0.3cm} 5043 MeV
\end{align}

 \begin{equation}
     b_{uu} = \hspace{0.3cm} b_{ud} = \hspace{0.3cm} b_{dd} = 112.46MeV
 \end{equation}

 \begin{equation}
     b_{us} = \hspace{0.3cm} b_{ds} \hspace{0.3cm}=\bigg(\frac{m_u}{m_s}\bigg)b_{uu} = 75.5 MeV
 \end{equation}

\begin{equation}
     b_{ss} =\bigg(\frac{m_u}{m_s}\bigg)b_{us}= \hspace{0.2cm} 50.72 MeV
 \end{equation}
 For charm sector,
 \begin{equation}
     b_{uc} = \hspace{0.3cm} b_{dc} = \bigg(\frac{m_u}{m_c}\bigg)b_{uu} = \hspace{0.3cm}23.8 MeV
 \end{equation}
\begin{equation}
     b_{sc} =\bigg(\frac{m_u}{m_c}\bigg)b_{us}= 15.98 MeV
\end{equation}
\begin{equation}
     b_{cc} = \bigg(\frac{m_u}{m_c}\bigg)b_{uc}\hspace{0.3cm} = 5.03 MeV
 \end{equation}
 In bottom sector,
\begin{equation}
    b_{ub} =  \hspace{0.3cm} b_{db}  = \bigg(\frac{m_u}{m_b}\bigg)b_{uu}= \hspace{0.3cm}8.07 MeV
\end{equation}
 \begin{equation}
     b_{sb} = \bigg(\frac{m_u}{m_s}\bigg)b_{ub}\hspace{0.3cm} =5.41MeV
 \end{equation}
 \begin{equation}
      \hspace{0.3cm} b_{cb} =\bigg(\frac{m_u}{m_c}\bigg)b_{ub}  \hspace{0.3cm} =2.73MeV,\hspace{0.3cm}
 \end{equation}
 \begin{equation}
     b_{bb} =\bigg(\frac{m_u}{m_b}\bigg)b_{ub} \hspace{0.3cm} = 0.579 MeV
 \end{equation}
By using the above values of parameters and quark masses, we can compute the effective quark masses for different spin-parity values.
\begin{table*}[ht!]
 \centering
\caption{Effective quark masses of the fully charmed, bottom, and charmed-bottom sector using an effective mass scheme where c stands for charm quark and b for bottom quark (in MeV).}c
\begin{tabular}{cccccccc}
       \hline
       \hline
Quark Content  & \hspace{0.3cm} Spin & \hspace{0.3cm} $m_{c}^{eff}$ & \hspace{0.3cm} $m_{\bar{c}}^{eff}$ & \hspace{0.3cm} $m_{b}^{eff}$ & \hspace{0.3cm} $m_{\bar{b}}^{eff}$ & \hspace{0.3cm} & \hspace{0.3cm} \\
         \hline
         
      $cccc\Bar{c}$  & \hspace{0.3cm} 5/2  & \hspace{0.3cm} 1712.5 & \hspace{0.3cm} 1712.5 & \hspace{0.3cm} - & \hspace{0.3cm} -
      \\
        $cccc\Bar{c}$  & \hspace{0.3cm} 3/2  & \hspace{0.3cm} 1710.6 & \hspace{0.3cm} 1704.9 & \hspace{0.3cm} - & \hspace{0.3cm} -
        \\
        $cccc\Bar{c}$  & \hspace{0.3cm} 1/2  & \hspace{0.3cm} 1708.7 & \hspace{0.3cm} 1705.6 & \hspace{0.3cm} - & \hspace{0.3cm}
       
         \\
         \hline
       
      $cccc\Bar{b}$  & \hspace{0.3cm} 5/2  & \hspace{0.3cm} 1712.2 & \hspace{0.3cm} 1712.2 & \hspace{0.3cm} - & \hspace{0.3cm} 5044.3
      \\
        $cccc\Bar{b}$  & \hspace{0.3cm} 3/2  & \hspace{0.3cm} 1711.2 & \hspace{0.3cm} 1711.2 & \hspace{0.3cm} -  & \hspace{0.3cm} 5040.2
        \\
        $cccc\Bar{b}$  & \hspace{0.3cm} 1/2  & \hspace{0.3cm} 1709.3 & \hspace{0.3cm} - & \hspace{0.3cm} - & \hspace{0.3cm} 5040.6
        \\
         \\
         \hline
       
      $cccb\Bar{b}$  & \hspace{0.3cm} 5/2  & \hspace{0.3cm} 1711.9 & \hspace{0.3cm} - & \hspace{0.3cm} 5044.1 & \hspace{0.3cm} 5044.1
      \\
        $cccb\Bar{b}$  & \hspace{0.3cm} 3/2  & \hspace{0.3cm} 1710.9 & \hspace{0.3cm} - & \hspace{0.3cm} 5043.8 & \hspace{0.3cm} 5040.8
        \\
        $cccb\Bar{b}$  & \hspace{0.3cm} 1/2  & \hspace{0.3cm} 1709.8 & \hspace{0.3cm} - & \hspace{0.3cm} 5040.8 & \hspace{0.3cm} 5040.8
        \\ \\
         \hline
       
      $ccbb\Bar{b}$  & \hspace{0.3cm} 5/2  & \hspace{0.3cm} 1711.6 & \hspace{0.3cm} - & \hspace{0.3cm} 5043.8 & \hspace{0.3cm} 5043.8
      \\
        $ccbb\Bar{b}$  & \hspace{0.3cm} 3/2  & \hspace{0.3cm} 1710.6 & \hspace{0.3cm} - & \hspace{0.3cm} 5043.6  & \hspace{0.3cm} 5041.3
        \\
        $ccbb\Bar{b}$  & \hspace{0.3cm} 1/2  & \hspace{0.3cm} 1709.6 & \hspace{0.3cm} - & \hspace{0.3cm} 5043.3 & \hspace{0.3cm} 5041.4
        \\ \\
         \hline
       
      $cbbb\Bar{b}$  & \hspace{0.3cm} 5/2  & \hspace{0.3cm} 1711.6 & \hspace{0.3cm} - & \hspace{0.3cm} 5043.5 & \hspace{0.3cm} 5043.5
      \\
        $cbbb\Bar{b}$  & \hspace{0.3cm} 3/2  & \hspace{0.3cm} 1710.3 & \hspace{0.3cm} - & \hspace{0.3cm} 5043.3  & \hspace{0.3cm} 5041.8
        \\
        $cbbb\Bar{b}$  & \hspace{0.3cm} 1/2  & \hspace{0.3cm} 1709.3 & \hspace{0.3cm} - & \hspace{0.3cm} 5043.1 & \hspace{0.3cm} 5041.9
        \\ \\
         \hline
       
      $bbbb\Bar{b}$  & \hspace{0.3cm} 5/2  & \hspace{0.3cm} - & \hspace{0.3cm} - & \hspace{0.3cm} 5043.2 & \hspace{0.3cm} 5043.2
      \\
        $bbbb\Bar{b}$  & \hspace{0.3cm} 3/2  & \hspace{0.3cm} - & \hspace{0.3cm} - & \hspace{0.3cm} 5043.1  & \hspace{0.3cm} 5042.4
        \\
        $bbbb\Bar{b}$  & \hspace{0.3cm} 1/2  & \hspace{0.3cm} - & \hspace{0.3cm} - & \hspace{0.3cm} 5042.8 & \hspace{0.3cm} 5042.4
        \\
       \hline
       \hline
       \end{tabular}
        \label{tab:3}
   \end{table*}

\subsection{Screened charge scheme }
The interaction among neighboring quarks modifies the quark masses, suggesting that the charge of the quark within the pentaquark might undergo similar effects. There is a linear dependency of effective charge on the charge of shielding quarks. Effective charge of a quark in exotic baryon Z$(a, b, c, d, f)$ is defined as:
\begin{equation}
    e_a^Z = e_a + \alpha_{ab} e_b + \alpha_{ac} e_c + \alpha_{ad} e_d + \alpha_{af} e_f
\end{equation}
\begin{equation}
    e_b^Z = e_b + \alpha_{ba} e_a + \alpha_{bc} e_c + \alpha_{bd} e_d + \alpha_{bf} e_f
\end{equation}
\begin{equation}
    e_c^Z = e_c + \alpha_{ca} e_a + \alpha_{cb} e_b + \alpha_{cd} e_d + \alpha_{cf} e_f
\end{equation}
\begin{equation}
    e_d^Z = e_d + \alpha_{da} e_a + \alpha_{db} e_b + \alpha_{dc} e_c + \alpha_{df} e_f
\end{equation}
\begin{equation}
    e_f^Z = e_f + \alpha_{fa} e_a + \alpha_{fb} e_b + \alpha_{fc} e_c + \alpha_{fd} e_d
\end{equation}
where $e_a$, $e_b$, $e_c$, $e_d$ and $e_f$  are the bare quark charges respectively. By considering the isospin symmetry, we take
\begin{align}
 \alpha_{ab} = \alpha_{ba}, \hspace{0.3cm} \alpha_{uu} =  \alpha_{ud} = \alpha_{dd} = \alpha, \hspace{0.3cm}  
\end{align}
\begin{align}
    \alpha_{us} = \alpha_{ds} = \alpha^{'},\hspace{0.3cm}
    \alpha_{ss} = \gamma
\end{align}
 For the charm sector:
\begin{align}
    \alpha_{uc} = \alpha_{dc} = \beta, \hspace{0.3cm} \alpha_{sc} = \delta^{'}, \hspace{0.3cm} \alpha_{cc} = \gamma^{'}
\end{align}
In a similar manner for the bottom sector:
\begin{align}
    \alpha_{ub} = \alpha_{db} = \beta^{''}, \hspace{0.3cm} \alpha_{sb} = \delta^{''}, \hspace{0.3cm} \alpha_{cb} = \gamma^{''}, \hspace{0.3cm} \alpha_{bb} = \xi
\end{align}
By using SU(3) flavor symmetry, we can further reduce these parameters, as given below:
\begin{align}
\alpha = \alpha{'} = \gamma,\hspace{0.7cm} \beta = \delta^{'}
\end{align}
Therefore, by using higher-order symmetries screening parameters can be converted into a single parameter. These parameters $\alpha_{ij}$ can be calculated as follows:
\begin{equation}
    \alpha_{ij} = \mid{\frac{m_i - m_j}{m_i + m_j}}\mid \times \delta
\end{equation}
where $\delta$ is taken as 0.81 as input parameter \cite{bains}.
\section{Magnetic moments in effective mass and screened charge scheme}
In this section, we study the magnetic moments of a fully heavy pentaquark system in all possible configurations. The magnetic moments of multiquark states provide valuable information about their internal structure. In a mutiquark system, the magnetic moment consists of two parts, as given below \cite{ortiz}:
\begin{equation}
    \mu = \mu_{spin}+\mu_{orbital}
\end{equation}
Since there is no orbital excitation, the magnetic moment depends only on the spin part.
\begin{equation}
    \mu_{spin} = \mu_s = \sum_i\mu^{eff}{\sigma_i}
\end{equation}
where,
\begin{equation}
    \mu^{eff} = \sum_{i}\frac{e^Z_i}{2m_i^{eff}}
\end{equation}
  By using the spin-flavor wave function of pentaquarks, the magnetic moments can be determined from the expectation value of equation (54) as follows:
 \begin{equation}
     \mu = \bra{\psi^{sf}}\mu_{s}\ket{\psi^{sf}}
 \end{equation}
 Here, $\psi^{sf}$ denotes the spin-flavor part of the pentaquark wavefunction and can be expressed as a product of flavor and spin wave function for different multiplets as:
 \begin{equation}
     \psi^{sf} = \phi_{f} \otimes \chi_{s}
 \end{equation}
 Now, we define the flavor and spin wavefunctions for different spin-parities. For $J^P$ = 5/2, where four quarks are identical its flavor wave function becomes:
 \begin{equation}
     \phi_{f} = \frac{1}{\sqrt{5}}[\alpha\alpha\alpha\alpha\bar{\beta}+\bar{\beta}\alpha\alpha\alpha\alpha+\alpha\bar{\beta}\alpha\alpha\alpha+\alpha\alpha\bar{\beta}\alpha\alpha+\alpha\alpha\alpha\bar{\beta}\alpha
     ]
 \end{equation}
Similarly, we can define the flavor wavefunctions for other pentaquark subsystems where two or three quarks are identical. \\
The spin wave function for spin 5/2 can be written as:
\begin{equation}
    \chi^{\frac{5}{2}}_s = \ket{\uparrow\uparrow\uparrow\uparrow\uparrow}
\end{equation}
By substituting the spin flavor wavefunction in the average value of the magnetic moment operator, we obtained the following expression:
\begin{equation}
    \mu_{\frac{5}{2}} = \mu_{\alpha}+\mu_{\alpha}+\mu_{\alpha}+\mu_{\alpha}+\mu_{\bar{\beta}}
\end{equation}
Similarly, for $J^P$ = 3/2 and  1/2 flavor wavefunction will remain the same while the spin part becomes different corresponding to their spin multiplet. The spin wavefunction for 3/2 is given as:
\begin{align}
      \chi^{\frac{3}{2}}_{s} = \frac{1}{2\sqrt{5}}\ket{4\uparrow\uparrow\uparrow\uparrow\downarrow-(\uparrow\uparrow\uparrow\downarrow+\uparrow\uparrow\downarrow\uparrow+\uparrow\downarrow\uparrow\uparrow+\downarrow\uparrow\uparrow\uparrow)\uparrow}
\end{align}
   
\begin{equation}
    \chi^{\frac{1}{2}}_{s}=\frac{1}{2\sqrt{3}}\ket{2(\uparrow\downarrow\uparrow\uparrow\downarrow-\downarrow\uparrow\uparrow\uparrow\downarrow)-(\uparrow\downarrow\uparrow\downarrow\uparrow+\uparrow\downarrow\downarrow\uparrow\uparrow+  \\ \nonumber \downarrow\uparrow\downarrow\uparrow\uparrow -\downarrow\uparrow\uparrow\downarrow\uparrow)}
\end{equation}
 
The direct product of flavor wavefunction defined in equation (55) and spin wavefunction in equations (58),(59) leads to the following expressions of the magnetic moment for 3/2 and 1/2 spin multiplets:
\begin{equation}
    \mu_{\frac{3}{2}} = \frac{9}{10}(\mu_{\alpha} +\mu_{\alpha} +\mu_{\alpha}+\mu_{\alpha})-\frac{3}{5}\mu_{\bar{\beta}}
\end{equation}
\begin{equation}
     \mu_{\frac{1}{2}} = \frac{2}{3}(\mu_{\alpha} +\mu_{\alpha})-\frac{1}{3}\mu_{\bar{\beta}}
\end{equation}
Thus, by using the above expressions, we computed the magnetic moments for full charm, bottom, and bottom-charm pentaquarks corresponding to their spin multiplets are shown in Table  3, and the obtained numerical results are compared with the available theoretical data.
\section{RESULTS AND DISCUSSION}
\subsection{Masses of fully heavy pentaquark states}

In the present work, we studied the masses of low-lying $S$ -Wave fully charm, bottom, and bottom-charm pentaquarks in the effective mass framework. The quantum numbers of these  states are  $J^P$= $\frac{1}{2}^{\pm}$,$\frac{3}{2}^{\pm}$ and $\frac{5}{2}^{\pm}$ respectively.
Also, we used input parameters given in the equations (16-30) to calculate the effective mass of each quark, which results from its neighboring quarks mediated by a single gluon exchange. Further, these masses helped us to predict the magnetic moments of fully heavy pentaquarks as described in the next section. We sum these effective quark masses given in table{3}to obtain the masses of fully heavy pentaquarks as mentioned in column 2 of Table {2}.
The numerical results obtained from the effective quark mass scheme predict that masses of fully heavy pentaquarks lie in the range of 8.537 MeV - 25212.8 MeV.
Therefore, our calculated results for masses of the fully charm $P_{cccc\bar{c}}$ and the fully bottom $P_{bbbb\bar{b}}$ shows reasonable agreement with Chiral quark model(ChQM) \cite{QuarkM} and Quark delocalization color screening model(QDCSM) \cite{QuarkM}. The comparison of our computed results listed in Table {2} also includes predictions based on the Lattice QCD-inspired quark model in the work of reference \cite{Gang}. The numerical results obtained from an effective quark mass scheme are compared with other works via the Constituent quark model \cite{CQM} and the prediction of charmed-bottom pentaquarks in the quark model. Furthermore, our results are compared with the MIT Bag model \cite{MIT}, the Chromomagnetic interaction (CMI) model \cite{chromo}, and QCD sum rules \cite{QCD}. Our predicted masses are in good agreement with the MIT Bag model and other works.
 In our work, we examine the discrepancies in mass between different spin multiplets. From Table {2}, we can see that the mass difference between $J^P$=$\frac{3}{2}$ and  $J^P$=$\frac{5}{2}$ in the fully charm sector is around 10-20 MeV while in the case of the bottom sector, there is a difference of 2 MeV. As the number of heavy quarks(N) increases, there is a substantial reduction in the mass difference of their spin multiplets($J$). Thus, the predicted results from our models would be helpful to understand the inner structure of the fully heavy pentaquark system in future studies.\\

\subsection{Magnetic moments of fully heavy   pentaquark states}
 Till now, no experimental information has been available for the magnetic moments of fully heavy pentaquark states. We aim to investigate them theoretically because they may be measured experimentally in the near future. In this study, we used the formalism described in the reference \cite{dhir} for baryons similar to exotic baryons (pentaquarks). By utilizing the Young tableau technique, we studied the pentaquark wave function and various spin symmetries. By substituting the magnetic moment operator between the pentaquark spin-flavor wave function, we obtained the expressions of magnetic moments corresponding to their spin symmetries listed in equations for fully heavy pentaquark systems. There is one spin
symmetry for spin-5/2 pentaquarks, four spin symmetries for spin-3/2 pentaquarks, and five spin symmetries
for spin-1/2 pentaquarks. The magnetic moments of fully heavy pentaquarks are computed by analyzing the relevant symmetries for possible spin configurations. For the calculation of magnetic moment, we used various input parameters like quark masses $m_i$ (where $i=u,d,s,c,b$), effective quark masses, and hyperfine terms $b_{ij}$ resulting from two-body interaction inside a pentaquark. Further, we compute the magnetic moments of fully heavy pentaquarks with quantum numbers  $J^P$= $\frac{1}{2}^{\pm}$,$\frac{3}{2}^{\pm}$ and $\frac{5}{2}^{\pm}$  within effective quark mass and screened charge framework.  In the effective mass scheme, we calculated numerical values for magnetic moments, and these values are generally smaller in magnitude compared to those obtained using the screened charge scheme. However, when we consider the influence of charge shielding, these values rise, opening up possibilities for future experimental investigations. The obtained results for the magnetic moments of fully heavy pentaquark states are mentioned in Table {4}. Also, we present a comparison of our numerical results with different theoretical models given in reference. Although these models show small variations with our predicted results they are still in qualitative agreement with the MIT Bag model for fully exotic pentaquarks in the work of reference \cite{MIT}.
 Our predicted results for the magnetic moments allow us to investigate their underlying dynamics and alternative pentaquark system topologies. It will be extremely fascinating to compare our results for both the masses and magnetic moments of fully heavy exotic pentaquarks with probable future results in effective quark mass and screened charge framework.
 \\
\section{SUMMARY}
 In 2020, the discovery of fully charmed tetraquark state X(6900) through LHCb observations sparked a keen interest in the potential existence of fully heavy pentaquark states \cite{X(6900)}.
Inspired by the experimental verification of $P_c$(4459) state, an enormous amount of theoretical efforts has begun to explore the inner dynamics of these states.
    In this work, we investigate the low- lying s -Wave fully heavy pentaquarks in a systematic manner $QQQQ\bar{Q}$ with spin-parity $J^P$= $\frac{1}{2}^{\pm}$,$\frac{3}{2}^{\pm}$ and $\frac{5}{2}^{\pm}$  using an effective quark mass and screened charge scheme. We first construct the color-spin wave function of fully heavy pentaquarks based on SU(2) and SU(3) symmetries. Then, by using the Young tableau technique, we obtained various spin symmetries corresponding to their spin multiplets in terms of Young-Yamuanouchi bases. After that, we calculated the masses and magnetic moments of $cccc\bar{c}$, $bbbb\Bar{b}$, and $cccb\Bar{b}$ or $bbbc\Bar{c}$ pentaquarks in their ground state configurations. \\
 The magnetic moment is strongly co-related with charge distributions of quarks inside a hadronic system. From our obtained results listed in Table 4, we can infer that the effect of shielding of charges increases the magnitude of the magnetic moment. The numerical results also indicate that magnetic moments vary with different configurations of pentaquarks. The magnetic moment of pentaquark states offers significant insights into the structure and size of hadrons. This analysis plays an important role in understanding the properties of hadrons.\\
In conclusion, the spectroscopy of fully heavy pentaquarks helps us to interpret the theoretical models of baryons and their structure. Our predicted results for both the masses and magnetic moments of pentaquarks can be investigated using other approaches such as the chiral quark model \cite{QuarkM}, Quark delocalized color screening model \cite{QuarkM}, the MIT bag model \cite{MIT}, the Lattice QCD inspired quark model \cite{Gang}, QCD sum rules \cite{QCD}, and the constituent quark model \cite{CQM}. It is suggested that these states could be investigated experimentally within $\Omega_{ccc}$ $J/\psi$ and $\Omega_{bbb}$ $\Xi_{bb}$ invariant mass spectrums.
 We expect that our predictions may provide valuable information for future experiments in their search for fully heavy pentaquarks as discussed in this work.

 \begin{table*}[ht]
 \centering
\caption{Magnetic moments of ground state fully heavy pentaquark states using the effective mass, screened charge, and
effective mass plus screened charge scheme together for $J^P$= $\frac{1}{2}^{\pm}$,$\frac{3}{2}^{\pm}$ and $\frac{5}{2}^{\pm}$ respectively. Magnetic
moments are in the unit of $\mu_N$.}
\begin{tabular}{cccccccc}
       \hline
       \hline
Quark Content & \hspace{0.3cm} S  & \hspace{0.3cm} Screened charge &  \hspace{0.3cm} Screened charge + effective mass &  \hspace{0.3cm} Effective mass & \hspace{0.3cm} Ref. \cite{MIT} & \hspace{0.3cm}  \\
         \hline
         \\
      $cccc\Bar{c}$  & \hspace{0.3cm} 1/2 & \hspace{0.3cm} 0.610  & \hspace{0.3cm} 0.611 & \hspace{0.3cm} 0.611 & \hspace{0.3cm} 0.83 \\
      \\
        $cccc\Bar{c}$ & \hspace{0.3cm} 3/2  & \hspace{0.3cm} 0.366 & \hspace{0.3cm} 0.589 & \hspace{0.3cm}0.365 & \hspace{0.3cm} 0.50 &  \\
        \\
          $cccc\Bar{c}$ & \hspace{0.3cm} 5/2  & \hspace{0.3cm} 1.09 &  \hspace{0.3cm} 1.10  & \hspace{0.3cm} 1.10 & \hspace{0.3cm} $\cdots$\\
          \\
           \hline
           \\
           $bccc\Bar{c}$  & \hspace{0.3cm} 1/2 & \hspace{0.3cm} 0.423 & \hspace{0.3cm} 0.424& \hspace{0.3cm} 0.420 & \hspace{0.3cm} 0.43 \\
           \\
        $bccc\Bar{c}$ & \hspace{0.3cm} 3/2  & \hspace{0.3cm} 0.079 & \hspace{0.3cm} -0.00022 & \hspace{0.3cm} 0.0006 & \hspace{0.3cm} 0.59 &  \\
        \\
          $bccc\Bar{c}$ & \hspace{0.3cm} 5/2  & \hspace{0.3cm} 0.477 &  \hspace{0.3cm}0.476  & \hspace{0.3cm} 0.669& \hspace{0.3cm} 0.90 \\
          \\
           \hline
           \\
           $bbcc\Bar{c}$  & \hspace{0.3cm} 1/2 & \hspace{0.3cm}0.257  & \hspace{0.3cm} 0.258 & \hspace{0.3cm} 0.230 & \hspace{0.3cm} 0.26 \\
      \\
        $bbcc\Bar{c}$ & \hspace{0.3cm} 3/2  & \hspace{0.3cm}-0.333 & \hspace{0.3cm}-0.334& \hspace{0.3cm}-0.205& \hspace{0.3cm} -0.01 &  \\
        \\
          $bbcc\Bar{c}$ & \hspace{0.3cm} 5/2  & \hspace{0.3cm} -0.0971 &  \hspace{0.3cm} -0.097& \hspace{0.3cm} 0.242& \hspace{0.3cm} 0.32\\
          \\
           \hline
           \\
           $bbbc\Bar{c}$  & \hspace{0.3cm} 1/2 & \hspace{0.3cm}0.186  & \hspace{0.3cm}0.187 & \hspace{0.3cm}0.164 & \hspace{0.3cm} 0.16 \\
      \\
        $bbbc\Bar{c}$ & \hspace{0.3cm} 3/2  & \hspace{0.3cm}-0.427 & \hspace{0.3cm}-0.428 & \hspace{0.3cm}-0.062 & \hspace{0.3cm} -0.30 &  \\
        \\
          $bbbc\Bar{c}$ & \hspace{0.3cm} 5/2  & \hspace{0.3cm}-0.624 &  \hspace{0.3cm} -0.624  & \hspace{0.3cm} -0.186& \hspace{0.3cm} -0.26 \\
          \\
           \hline
          \\ $bbbb\Bar{c}$  & \hspace{0.3cm} 1/2 & \hspace{0.3cm}0.070  & \hspace{0.3cm}0.071 & \hspace{0.3cm}0.039& \hspace{0.3cm} 0.05 \\
      \\
        $bbbb\Bar{c}$ & \hspace{0.3cm} 3/2  & \hspace{0.3cm}-0.753& \hspace{0.3cm}-0.754 & \hspace{0.3cm}-0.419 & \hspace{0.3cm} -0.65 &  \\
        \\
          $bbbb\Bar{c}$ & \hspace{0.3cm} 5/2  & \hspace{0.3cm} -1.104 &  \hspace{0.3cm} -1.104  & \hspace{0.3cm}-0.614& \hspace{0.3cm} $\cdots$\\
          \\
           \hline
         \\
          $cccc\Bar{b}$ & \hspace{0.3cm} 1/2  & \hspace{0.3cm}0.498   & \hspace{0.3cm}0.499  &  \hspace{0.3cm}0.468& \hspace{0.3cm} 0.62 \\
          \\
           $cccc\Bar{b}$ & \hspace{0.3cm} 3/2  & \hspace{0.3cm}1.021 & \hspace{0.3cm}1.021&  \hspace{0.3cm}0.7224& \hspace{0.3cm} 1.08\\
          \\
            $cccc\Bar{b}$ & \hspace{0.3cm} 5/2  & \hspace{0.3cm}2.01 & \hspace{0.3cm}2.01 & \hspace{0.3cm} 1.52&\hspace{0.3cm}  1.47
            \\
             \hline
         
          \\
          $cccb\Bar{b}$ & \hspace{0.3cm} 1/2  & \hspace{0.3cm}0.232   & \hspace{0.3cm}0.417  &  \hspace{0.3cm} 0.182& \hspace{0.3cm} 0.62 \\
          \\
           $cccb\Bar{b}$ & \hspace{0.3cm} 3/2  & \hspace{0.3cm}0.613 & \hspace{0.3cm}0.614 &  \hspace{0.3cm} 0.365 & \hspace{0.3cm} 1.08\\
          \\
            $cccb\Bar{b}$ & \hspace{0.3cm} 5/2  & \hspace{0.3cm}1.39  & \hspace{0.3cm}1.39 & \hspace{0.3cm} 1.10 &\hspace{0.3cm}  1.47
            \\
             \hline
       
          \\
          $ccbb\Bar{b}$ & \hspace{0.3cm} 1/2  & \hspace{0.3cm} 0.223  & \hspace{0.3cm} 0.337 & \hspace{0.3cm} 0.277& \hspace{0.3cm} 0.37
          \\
          \\
           $ccbb\Bar{b}$ & \hspace{0.3cm} 3/2  & \hspace{0.3cm} 0.543  &  \hspace{0.3cm} 0.543  &  \hspace{0.3cm} 0.508  & \hspace{0.3cm} 0.54 &
          \\
          \\
            $ccbb\Bar{b}$ & \hspace{0.3cm} 5/2  & \hspace{0.3cm}0.821  & \hspace{0.3cm} 0.820  & \hspace{0.3cm} 0.67 & \hspace{0.3cm} 0.88 \\
           \hline

          \\
          $cbbb\Bar{b}$ & \hspace{0.3cm} 1/2  & \hspace{0.3cm} 0.049   & \hspace{0.3cm} 0.101  & \hspace{0.3cm} 0.086  & \hspace{0.3cm} 0.06
          \\
          \\
           $cbbb\Bar{b}$ & \hspace{0.3cm} 3/2  & \hspace{0.3cm} 0.225  & \hspace{0.3cm} 0.225 &  \hspace{0.3cm} 0.223 & \hspace{0.3cm} 0.35
          \\
          \\
            $cbbb\Bar{b}$ & \hspace{0.3cm} 5/2 & \hspace{0.3cm} 0.294  & \hspace{0.3cm} 0.294 &  \hspace{0.3cm} 0.241  &\hspace{0.3cm} 0.31
            \\
             \hline
          \\
          $bbbb\Bar{b}$ & \hspace{0.3cm} 1/2 & \hspace{0.3cm} -0.103 & \hspace{0.3cm} -0.103  &  \hspace{0.3cm} -0.103 & \hspace{0.3cm} -0.14 \\
          \\
           $bbbb\Bar{b}$ & \hspace{0.3cm} 3/2 & \hspace{0.3cm} -0.062   & \hspace{0.3cm} -0.062 &  \hspace{0.3cm} -0.062 & \hspace{0.3cm} -0.08 \\
          \\
            $bbbb\Bar{b}$ & \hspace{0.3cm} 5/2  & \hspace{0.3cm} -0.186  & \hspace{0.3cm} -0.186 &  \hspace{0.3cm} -0.186 & \hspace{0.3cm} $\cdots$\\
            \\
       
       \hline
       \hline
       \end{tabular}
        \label{tab:mm}
   \end{table*}

\nocite{*}

\bibliography{Fully}% Produces the bibliography via BibTeX.

\end{document}